\documentclass[runningheads]{llncs}
\usepackage{booktabs}
\usepackage{graphicx}
\usepackage{amsmath}
\begin{document}
\title{Simultaneous Bone and Shadow Segmentation Network using Task Correspondence Consistency}
\author{Aimon Rahman\textsuperscript{1}, Jeya Maria Jose Valanarasu\textsuperscript{1}, Ilker Hacihaliloglu\textsuperscript{2}, Vishal M Patel\textsuperscript{1}} 
\institute{Johns Hopkins University \and University of British Columbia}
\titlerunning{Simultaneous Bone and Shadow Segmentation Network}


%
\maketitle              % typeset the header of the contribution
\begin{abstract}
Segmenting both bone surface and the corresponding acoustic shadow are fundamental tasks in ultrasound (US) guided orthopedic procedures. However, these tasks are challenging due to minimal and blurred bone surface response in US images, cross-machine discrepancy, imaging artifacts, and low signal-to-noise ratio. Notably, bone shadows are caused by a significant acoustic impedance mismatch between the soft tissue and bone surfaces. To leverage this mutual information between these highly related tasks, we propose a single end-to-end network with a shared transformer-based encoder and task independent decoders for simultaneous bone and shadow segmentation. To share complementary features, we propose a cross task feature transfer block which learns to transfer meaningful features from decoder of shadow segmentation to that of bone segmentation and vice-versa. We also introduce a correspondence consistency loss which makes sure that network utilizes the inter-dependency between the bone surface and its corresponding shadow to refine the segmentation. Validation against expert annotations shows that the method outperforms the previous state-of-the-art for both bone surface and shadow segmentation.

\end{abstract}
\section{Introduction}

There has been a significant interest in incorporating ultrasound (US) imaging for computer assisted orthopedic surgery (CAOS) procedures owing to its non-invasive, radiation-free, and cost-effective nature. However, due to bone surfaces appearing only several millimeters (mm) in thickness along with noisy artifacts, researchers have been focusing on developing automated bone segmentation and enhancement methods \cite{hacihaliloglu2017ultrasound}. These bone surfaces generally have the highest intensity in US images which is then followed by a low-intensity region, namely bone shadows. Bone shadow is the result of a high acoustic impedance mismatch between the bone surface and the adjacent soft tissue, which reflects the US signal to the transducer. The bone shadow information is essential to guide the orthopedic surgeon to a standardized viewing plane with minimal noise and artifacts. Hence, both bone surface and shadow segmentation are crucial to CAOS procedures. 

Recent literature on bone and shadow segmentation focus on  learning individual networks for each problem separately  \cite{alsinan2019automatic,wang2020cnn,alsinan2020bone,alsinan2019spine}. However, in \cite{wang2018simultaneous}, Wang et al. \cite{wang2018simultaneous} proposed a pre-enhancement network that leverages bone shadow information for bone surface segmentation. The bone shadow was obtained using a bone shadow enhancement method where a signal transmission map is constructed from the local phase bone image features \cite{hacihaliloglu2017enhancement}. The enhanced bone shadow information has also been used in \cite{wang2020robust} where a multi-task learning-based method to segment bone shadow region is proposed.

 It should be noted that bone shadow is a signal void that indicates the loss of energy as US waves propagate through bone tissues. Thus, the quality of bone surface segmentation can have major impact on shadow segmentation accuracy and vice-versa. However, existing works do not fully exploit the structure of these highly related tasks. Despite being closely-related, existing top networks for bone and shadow segmentation have significantly different and specialized architectures. Our proposed method explores the idea of exploiting shared features for a more compact network and taking advantage of interactions between the two tasks to generate a better feature representation. We hypothesize that the interrelation between bone and shadow response in US images can be leveraged to significantly improve the quality of both learned networks. In summary, we present the following contributions in this paper:

\begin{itemize}
    \item We are the first to integrate two highly-related homogeneous tasks into a single framework for unified bone surface and shadow segmentation. The common encoder brings powerful synergy across both tasks when extracting shared deep features for the two tightly-coupled problems.
    
    \item We propose a cross task feature transfer block to extract complementary features at decoders to improve the quality of performance in the multi-task learning framework.
    
    \item We propose a task correspondence consistency loss to further regularize the network by ensuring the transitivity between the two related predictions.

    \item We conduct extensive experiments using the in vivo US scans of knee, femur, distal radius, spine, and tibia bones collected using two US machines and demonstrate that the proposed method is competitive with other individual specialized state-of-the-art methods.
    
\end{itemize}

\section{Method}

\subsection{Preliminaries}
\begin{figure}[!htb]
\centering
\vspace{-2.5 em}
\includegraphics[width=0.9\linewidth]{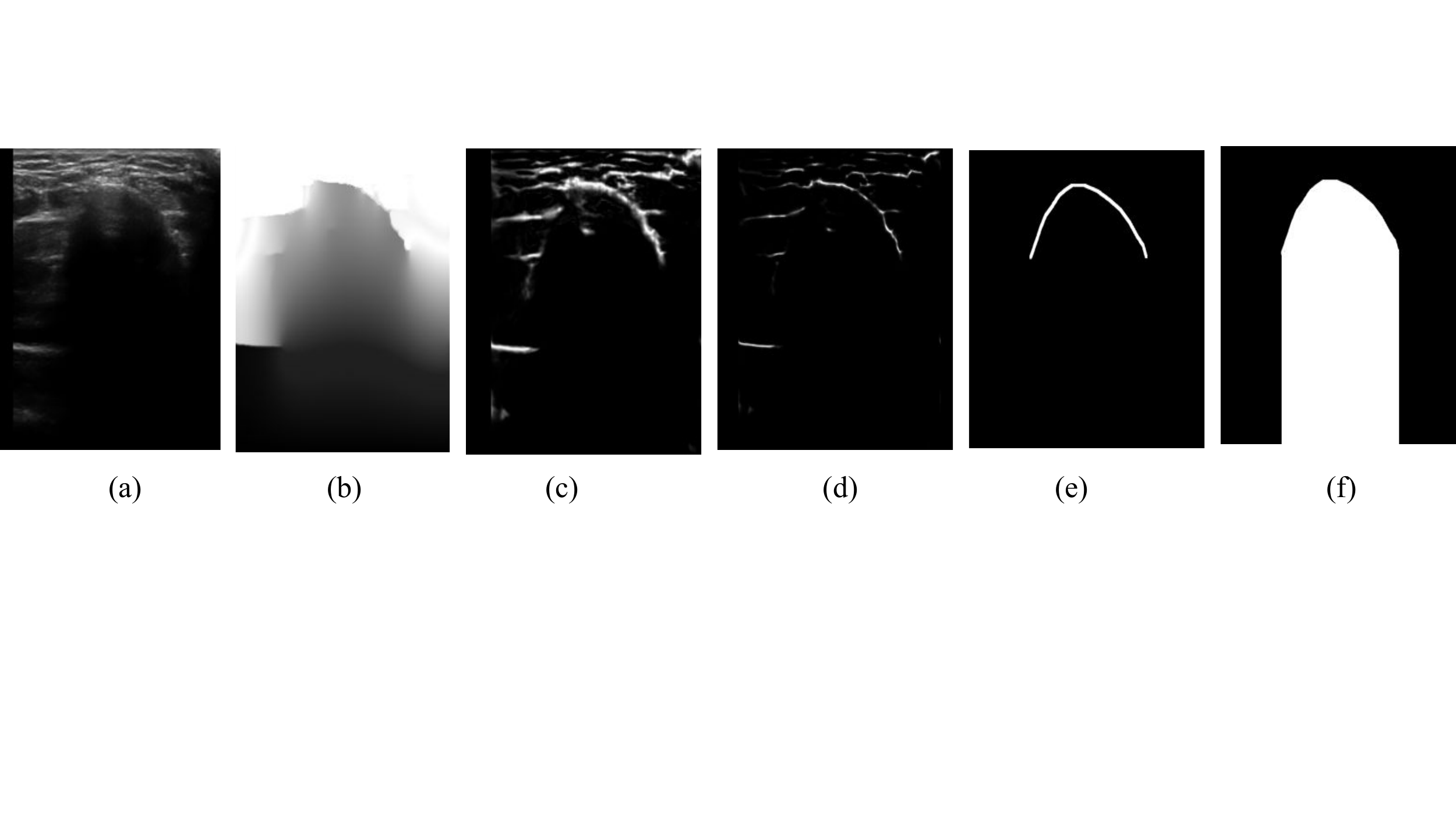}
\vspace{-8 em}
\caption{(a) B-mode US scan (b) LPT (c) LP (d) BSE (e)Bone Surface Segmentation and (f) Bone Shadow Segmentation.}
\label{Fig:example}
\end{figure}

 Instead of using only B-mode US scan as input, the proposed network takes the concatenation of three filtered images along with the original B-mode US scan $(US(x,y))$. The filtered images are shown in Fig. \ref{Fig:example}(a)-(d). This has been done to reduce the domain discrepancy between the images obtained using different US machine settings or different orientations of the transducer. During the extraction of filtered images we have used the original parameters and constant values described in \cite{hacihaliloglu2014local,hacihaliloglu2017enhancement}. The Local Phase Tensor Image $( L P T(x, y))$ is computed by defining odd and even filter responses using \cite{hacihaliloglu2014local}. Local Phase Bone Image $L P(x, y)$  is computed using: $L P(x, y)=L P T(x, y) \times L P E(x, y) \times \operatorname{LwP} A(x, y)$, where $L P E(x, y)$ and $L w P A(x, y)$ represent the local phase energy and local weighted mean phase angle image features, respectively. These two features are computed using monogenic signal theory as \cite{hacihaliloglu2017enhancement}. Bone Shadow Enhanced image $BSE(x,y)$ is obtained by modeling the interaction of Ultrasound signal at position (x,y) within the tissues as scattering and attenuation information using the method proposed in \cite{hacihaliloglu2017enhancement},
 \setlength{\belowdisplayskip}{0pt} \setlength{\belowdisplayshortskip}{0pt}
\setlength{\abovedisplayskip}{0pt} \setlength{\abovedisplayshortskip}{0pt}
\begin{equation*}
    BSE(x,y)=[(CM_{LP}(x,y)-\rho )/[max(US_{A}(x,y),\epsilon )]^\delta ]+\rho
\end{equation*}
Here the confidence map is denoted by $CM_{LP}(x,y)$ which is obtained by modeling the US signal propagation inside the tissue considering bone feature in local phase bone image LP(x, y). $US_{A}(x,y)$ maximizes the visibility of bone features with high intensity inside a local region. $\delta$ represents the tissue attenuation coefficient. $\rho$ is related to echogenicity confining the bone surface and $\epsilon$ is a small constant to avoid division by zero.

\vspace{-3.5 em}
\subsection{Network Architecture}
\begin{figure}[!htb]
\centering
\vspace{-1 em}
\includegraphics[page=2,width=1\linewidth]{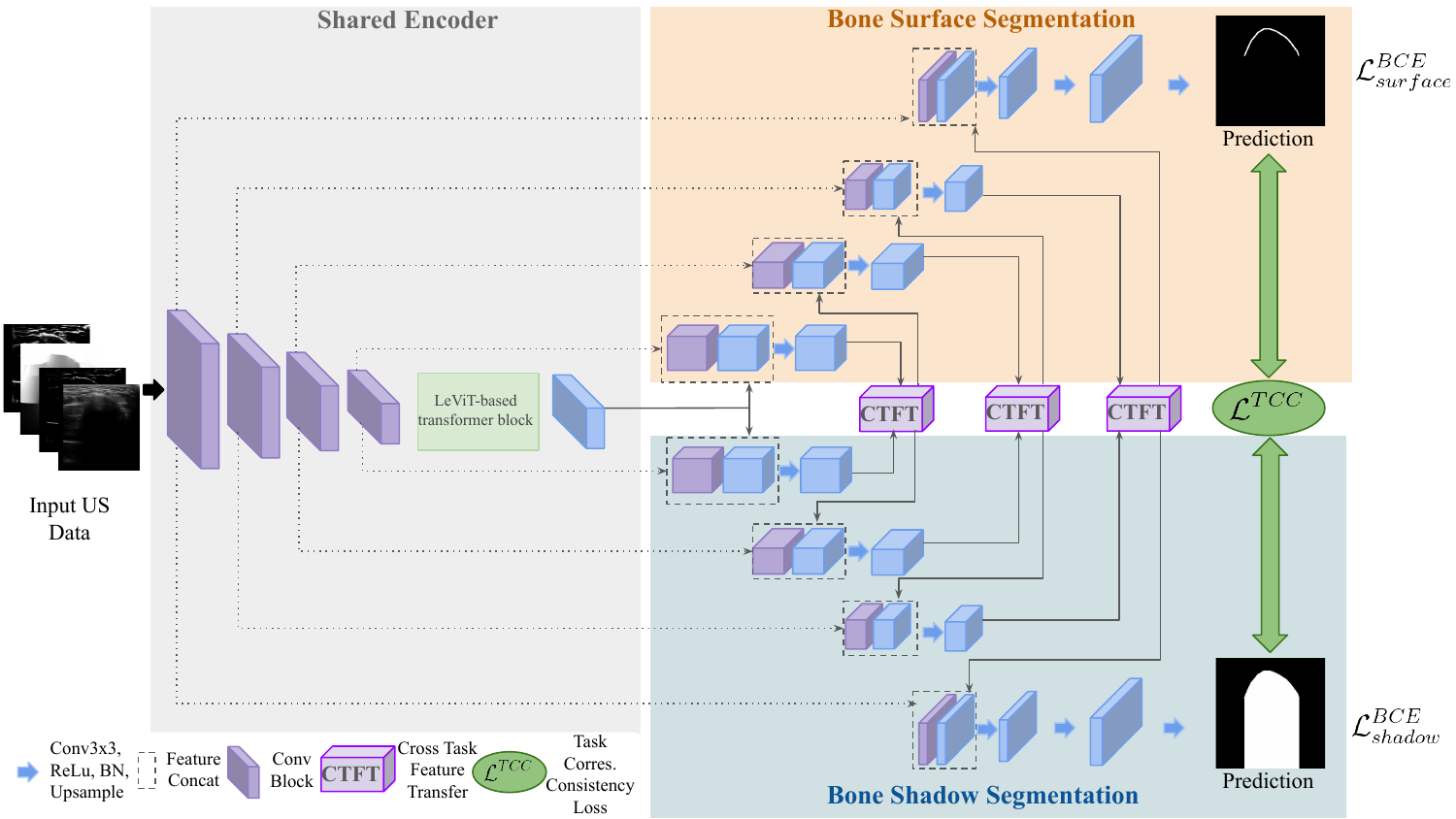}
\vspace{-2 em}
\caption{An overview of the proposed SSNet for simultaneous bone surface and shadow segmentation from US images.}
\label{Fig:network}
\end{figure}

We propose Shadow and Surface Segmentation Network (SSNet)  for simultaneous bone surface and shadow segmentation from US images which is illustrated in Figure \ref{Fig:network}. SSNet is composed of a shared LeViT-based encoder to extract global and long-range spatial features and two CNN-based decoders with a cross task feature transfer block to leverage mutual information between the two tasks.
\vspace{-1.5 em}
\subsubsection{(i) LeViT-based Shared Encoder:}
The shared encoder for bone and shadow surface segmentation is built based on the LeViT architecture \cite{graham2021levit}. The encoder part consists of four $3\times3$ convolution layers with stride 2 initially followed by three transformer blocks.  Features from the convolution layers are forwarded to the LeViT transformer blocks which require fewer floating-point operations (FLOPs) than ViTs \cite{dosovitskiy2020image}. 
The local and global features at different scales are exploited by concatenating the features from both transformer and convolution layers. 
\vspace{-1.5 em}

\subsubsection{(ii) CNN-based Decoders:}
The decoder part of the network consists of two separate branches for bone surface and shadow segmentation. Inspired by UNet \cite{ronneberger2015u}, the features from decoders are concatenated with skip connection to effectively reuse spatial information of feature maps. The resolution from the previous layers is recovered using the cascaded upsampling technique similar to UNet. The decoder blocks consist of a $3\times3$ convolution, batch normalization layer followed by a ReLU layer.

\subsection{Cross Task Feature Transfer Block}
\begin{figure}[htbp]
\vspace{-1 em}
\centering
\includegraphics[width=0.85\linewidth]{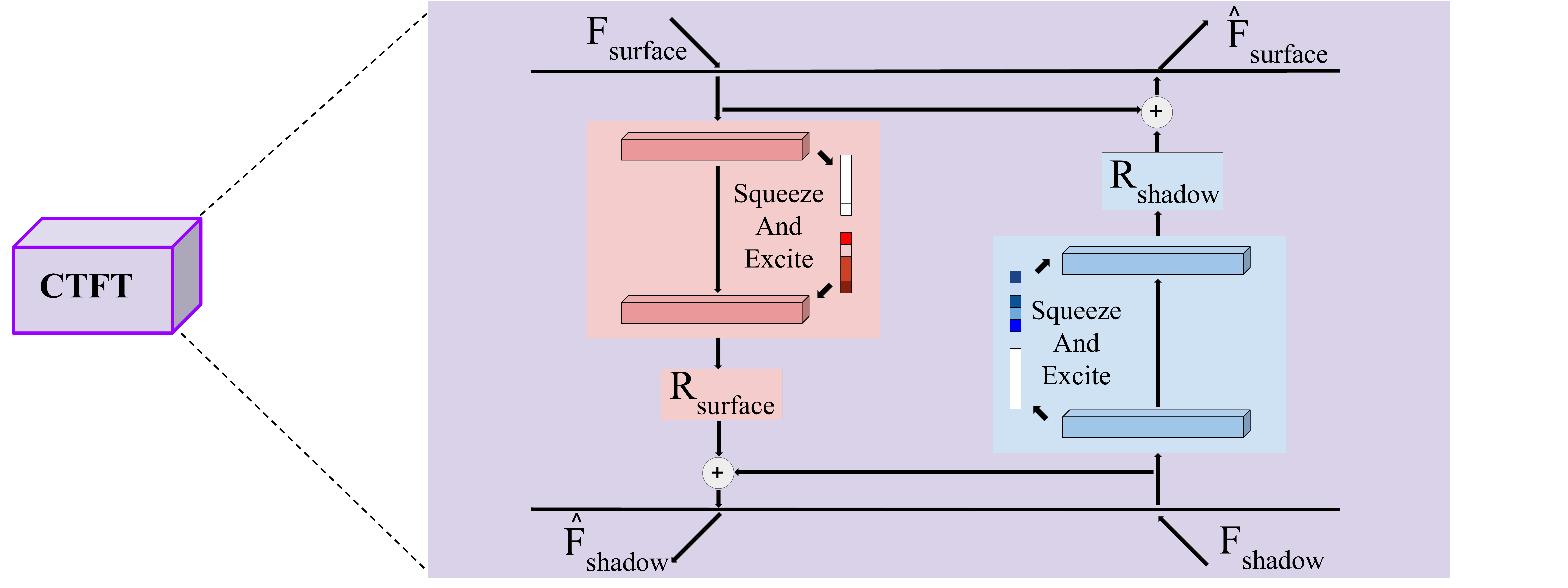}
\vspace{-1 em}
\caption{An overview of the Cross Task Feature Transfer Block.}
\label{Fig:details}
\end{figure}

\vspace{-1 em}
To leverage the joint-learning capabilities of these two highly-related tasks, we propose a cross task feature transfer (CTFT) block used in between the two decoders. CTFT extracts complementary features from the two decoder branches using a squeeze and excitation block \cite{hu2018squeeze} and forwards them to the next decoder blocks of respective branches. We use squeeze and excitation block to learn which features of the surface segmentation decoder would help in segmenting bone shadow and vice-versa. Squeeze and excite enables dynamic channel-wise feature re-calibration thus help extract features that contributes to the complementary task.  The details of CTFT are illustrated in Figure \ref{Fig:details}. It takes in two inputs: $F_{surface}$ and $F_{shadow}$ corresponding to the feature maps of bone surface and shadow decoders. $F_{surface}$ is passed through a squeeze and excite layer to obtain the residual $R_{surface}$ which is added to $F_{shadow}$ to obtain $\hat{F}_{shadow}$. $\hat{F}_{shadow}$ is then passed to the next block of the shadow segmentation decoder. Similarly, $F_{shadow}$ is passed through a squeeze and excite layer to obtain the residual $R_{shadow}$ which is added to $F_{surface}$ to obtain $\hat{F}_{surface}$.

\vspace{-1 em}
\subsection{Task Correspondence Consistency Loss} 
To guarantee both networks capture the inter-dependency between bone surface and its corresponding shadow, we introduce two additional loss terms called Task Correspondence Consistency Loss. For an US image $X \in \mathcal{X}$, the annotations $Y = \left(y_{1}, y_{2}\right)$  is a set of labels containing bone surface and shadow segmentation masks, respectively. Let, $\hat{Y} = \left(\hat{y_{1}}, \hat{y_{2}}\right)$ be the predictions of the decoder networks. Our additional loss term includes two mapping  $F_{1}: y_{1} \rightarrow y_{2}$ and $F_{2}: y_{2} \rightarrow y_{1}$. For any US image $X$, each loss term ensure consistency by translating in between bone surface and shadows, i.e., $y_1 \rightarrow F_1(y_1) \approx y_2$. The task corresponding consistency loss further regularizes the network to produce robust segmentation masks for both task and prevent them to contradict each other. The proposed Task Correspondence Consistency Loss $ \mathcal{L}^{TCC}(X, Y)$ is defined as:
\setlength{\belowdisplayskip}{0pt} \setlength{\belowdisplayshortskip}{0pt}
\setlength{\abovedisplayskip}{0pt} \setlength{\abovedisplayshortskip}{0pt}
\begin{equation*}
    \mathcal{L}^{TCC}(X, Y) =  \mathcal{L}^{BCE}(y_1,F_2(\hat{y_2})) +  \mathcal{L}^{BCE}(y_2,F_1(\hat{y_1})).
\end{equation*}

\vspace{-0.5cm}
\section{Experiments and Results}
\noindent \textbf{Dataset: } The study includes 25 healthy volunteers with the approval of the institutional review board (IRB). Total 1042 different US images have been collected using SonixTouch US machine (Analogic Corporation, Peabody, MA, USA) with 2D C5-2/60 curvilinear and L14-5 linear transducer. For independent testing, 3 new subjects have been included in the study. Using handheld wireless US scans (Clarius C3, ClariusMobile Health Corporation, BC, Canada), a total of 185 scans have been collected. Depending on the depth setting, scan resolution varies between 0.1mm to 0.15mm. As both transducer and reconstruction pipelines are different, Clarius have low image quality. The scans include knee, femur, radius, and spine data and all of them are manually segmented by an expert ultrasonographer. For the Sonix dataset, a random 80:20 split has been applied based on the subject, making the final training set with 834 samples and the test set with 208 samples.

\noindent \textbf{Implementation Details:} SSNet is trained using a batch size of 32. For training both branches, a two-step training phase is adapted. Each of these steps are trained until convergence. The weights and bias of the network are optimized using Adam optimizer with a learning rate of $10^{-4}$. All US scans and their corresponding masks are resized to $224 \times 224$ pixels and rescaled between 0 to 1. All transformer blocks in the LeViT architecture were pre-trained on ImageNet-1k. The overall loss function we use to train the multi-task network is,
\begin{equation*}
    \mathcal{L}^{total}(X, Y) = \mathcal{L}^{BCE}(y_1,\hat{y_1}) + \mathcal{L}^{BCE}(y_2,\hat{y_2}) + \mathcal{L}^{TCC}(X, Y).
\end{equation*}
Binary-cross entropy loss has been used between the prediction and the ground truth, which is expressed as,
\begin{equation*}
\mathcal{L}_{C E(p, \hat{p})}=-\left(\frac{1}{w h} \sum_{x=0}^{w-1} \sum_{y=0}^{h-1}(p(x, y) \log (\hat{p}(x, y)))+(1-p(x, y)) \log (1-\hat{p}(x, y))\right).
\end{equation*}
Here, $w$ and $h$ represents the dimension of ultrasound scan, $p(x, y)$ denotes the pixel in scan and $\hat{p}(x, y)$ denotes the output prediction at a specific location
$(x, y)$. Test images can be forwarded through the network for both tasks in one shot. The experiments are carried out on a Linux workstation with Intel 3.50 GHz CPU and a 12GB NVidia Titan Xp GPU using the PyTorch framework. Dice coefficients are used to measure the segmentation performance of different methods.

\noindent \textbf{Quantitative Comparison:} For bone shadow segmentation, we compare the performance of our proposed method with that of UNet \cite{ronneberger2015u}, MFG-CNN \cite{wang2018simultaneous}, and PSPGAN MTL \cite{wang2020robust}.  PSPGAN MTL is the current state-of-the-art for bone shadow segmentation. For bone surface segmentation, we compare with UNet \cite{ronneberger2015u}, MFG-CNN \cite{wang2018simultaneous} without the classification labels, and LPT+GCT \cite{wang2020cnn}.  All the methods are trained using the same training dataset as used to train the proposed method. PSPGAN-MTL uses a conditional shape discriminator to enforce bone interval boundaries which provides more accurate and robust bone segmentation. Instead of using bone interval boundaries during the training, we enforce the boundary from the bone surface segmentation mask during inference instead. Average test results are shown in Table \ref{Tab:results1}. It can be observed that the shared network SSNet outperforms the current state-of-the-art \cite{wang2020robust} and individual networks for both bone and shadow segmentation (paired t-test $<$ 0.05). 

\begin{table}[htbp]
\caption{Results averaged over 5 folds. Numbers correspond to dice score with  standard deviation. Boldface numbers indicate the best segmentation performance.}
\label{Tab:results1}
\vspace{-15pt}
\centering
\medskip
\resizebox{1\columnwidth}{!}{%
\begin{tabular}{@{}lcccccccccccc@{}}
\toprule
& \multicolumn{2}{c}{SonixTouch} & \multicolumn{2}{c}{Clarius} \\
\cmidrule(lr){2-3} \cmidrule(lr){4-5}
Method & Surface (\%) & Shadow (\%) & Surface (\%) & Shadow (\%) \\
\midrule
UNet \cite{ronneberger2015u} ~ &  $76.01\pm0.20$ & $88.33\pm0.06$ & $75.11\pm0.31$ & $84.03\pm0.14$ \\
MFG-CNN \cite{wang2018simultaneous} ~ &  $81.05\pm0.06$ & $-$ & $82.23\pm0.14$ & $-$ \\
LPT+GCT \cite{wang2020cnn} ~ & $81.65\pm0.10$& $-$ & $83.05\pm0.21$ & $-$ \\
PSPGAN-MTL \cite{wang2020robust} ~ & $-$ &  $93.49\pm0.06$ & $-$  & $91.01\pm0.18$  \\
% SSNet+ CTFT ~ &  $84.03\pm0.11$ & $94.88\pm0.16$ &  $81.13\pm0.19$ &  $92.43\pm0.18$ &  \\
SSNet+ CTFT + TCC loss (ours) ~ & $\mathbf{87.03\pm0.21}$ & $\mathbf{96.18\pm0.43}$ & $\mathbf{83.33\pm0.31}$  & $\mathbf{93.01\pm0.23}$ \\
\bottomrule
\hline
\end{tabular}
}
\end{table}

\begin{figure}[htbp]
\centering
\includegraphics[width=1\linewidth]{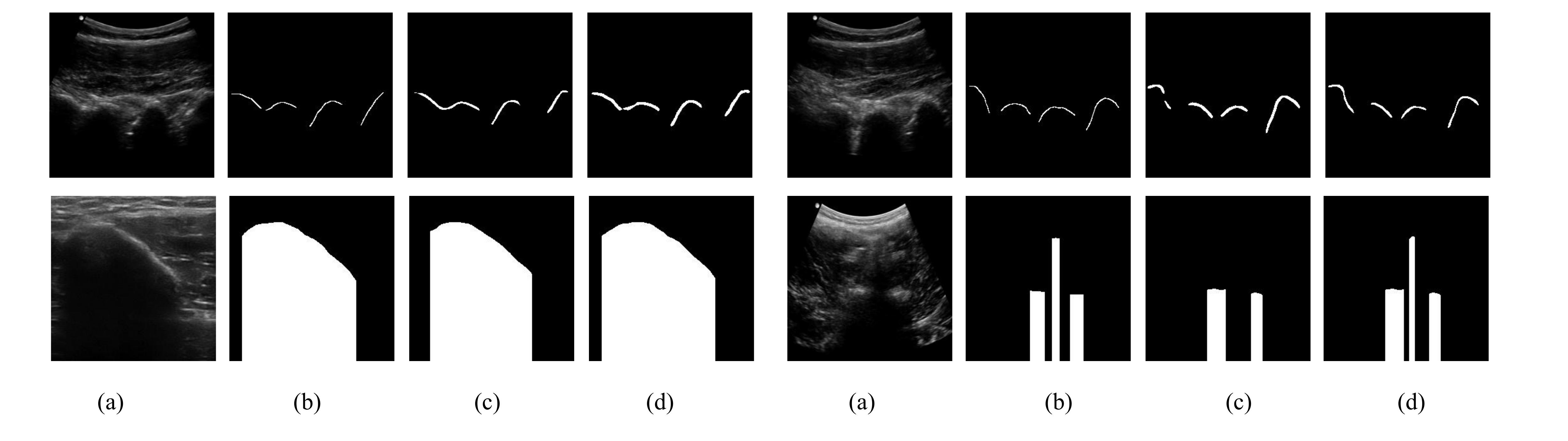}
\vskip-15pt\caption{ Top Row -  Bone surface segmentation. Bottom Row - Bone shadow segmentation. (a) Input US scan (b) Ground Truth (c) Output from current state-of-the-art \cite{wang2020cnn} (surface), \cite{wang2020robust} (shadow) (d) Ours.}
\label{Fig:sota}
\end{figure}

\noindent \textbf{Qualitative Comparison: } We present sample qualitative results in Fig. \ref{Fig:sota} for both bone surface and shadow segmentation. It can be observed that the current state-of-the-art methods result in either missed shadow regions or disjoint bone segmentation maps. As our proposed method uses the inter-dependency between these tasks, we see a significant improvement with less discrepancies compared to the ground truth annotations.

\vspace{-2 em}
\section{Discussion}
\noindent \textbf{Ablation Study: } To understand the contribution of each individual module in the proposed SSNet, we conduct an ablation study and report it in Table \ref{Tab:abl}. It can be observed that addition of CTFT helps improve the performance of both surface and shadow segmentation by injecting complementary features to the respective decoders. Also, using the propose task consistency ($\mathcal{L}^{TCC}$) further regularizes the network and boosts the segmentation performance.

\begin{table}[htbp]
\caption{Ablation Study. Numbers correspond to dice score.}
\label{Tab:abl}
\centering
\medskip
\resizebox{1\columnwidth}{!}{%
\begin{tabular}{@{}lcccccccccccc@{}}
\toprule
& \multicolumn{2}{c}{SonixTouch} & \multicolumn{2}{c}{Clarius} \\
\cmidrule(lr){2-3} \cmidrule(lr){4-5}
Method & Surface (\%) & Shadow (\%) & Surface (\%) & Shadow (\%) \\
\midrule
SSNet (Base)  ~ &  $82.95\pm0.13$ & $93.34\pm0.06$ & $81.71\pm0.20$ & $90.94\pm0.22$ \\
SSNet  + CTFT ~ &  $84.03\pm0.11$ & $94.88\pm0.16$ &  $81.13\pm0.19$ &  $92.43\pm0.18$ &  \\
SSNet + CTFT +  $\mathcal{L}^{TCC}$ (ours) ~ & $\mathbf{87.03\pm0.21}$ & $\mathbf{96.18\pm0.43}$ & $\mathbf{83.33\pm0.31}$  & $\mathbf{93.01\pm0.23}$ \\
\bottomrule
\hline
\end{tabular}
}
\end{table}

\begin{figure}[htbp]
\centering
\includegraphics[width=1\linewidth]{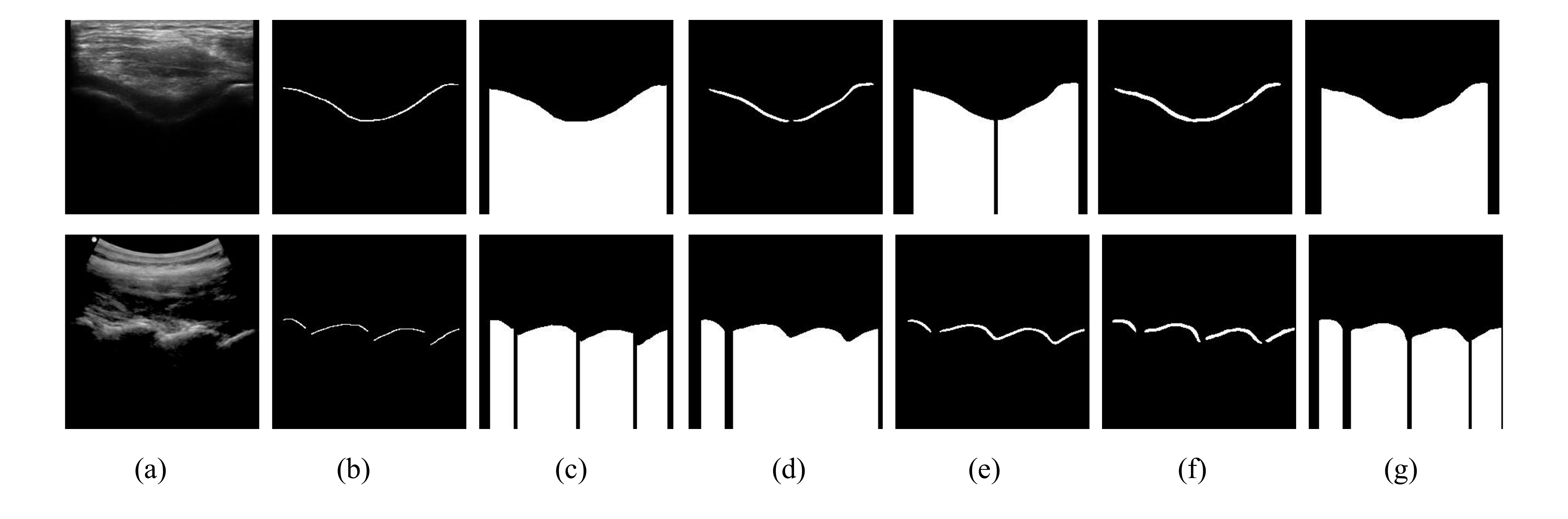}
\vskip-17pt\caption{(a) Input US scan (b) Surface ground truth (c) Shadow ground truth (d) Individual network output (e) Output generated from cascaded network (f) Surface output from ours (g) Shadow output from ours.}
\label{Fig:cascade_vs_ours}
\end{figure}

\noindent \textbf{Importance of joint learning: } Qualitative results in Fig \ref{Fig:cascade_vs_ours} shows the importance of the joint learning framework.  The result from cascaded network demonstrates that the faulty output from either of the network can produce wrong corresponding prediction. Cascaded network corresponds to using a deep network to predict the bone shadow map from bone surface segmentation map and vice-versa. For example, missing or joint boundaries in bone surface segmentation may result in wrong bone intervals in shadow network as demonstrated in the top row of Fig \ref{Fig:cascade_vs_ours}. Similarly, over or under-segmented bone shadow predictions may produce faulty surface estimations. However, as each of the decoders in our network is specialized for their respective task and further regularized by ensuring cross-task consistency, our network produces more consistent results.

%  \noindent \textbf{Analysis on LeViT architecture} 
% We explore the performance of different LeViT-based architectures. The number of channels fed into the first block of the transformer represents the type of encoders used in this experiment, namely LeViT-128s, LeViT-192, and LeViT-384. From Table \ref{Tab:results_levit} it can be observed that LeVit-384 yields the best results.
 
% \begin{table*}[!h]
% \caption{Ablation Study. All results are reported in Dice score.}
% \label{Tab:results_levit}
% %\setlength{\tabcolsep}{3.8pt}
% \centering
% \medskip
% \begin{tabular}{@{}lcccccccccccc@{}}
% \toprule
% & \multicolumn{2}{c}{SonixTouch} & \multicolumn{2}{c}{Clarius} \\
% \cmidrule(lr){2-3} \cmidrule(lr){4-5}
% Method & Surface (\%) & Shadow (\%) & Surface (\%) & Shadow (\%) \\
% \midrule
% LeViT-UNet-128s \cite{xu2021levit} ~ &  $78.75\pm0.17$ & $88.34\pm0.14$ & $77.94\pm0.24$ & $86.04\pm0.31$ \\
% LeViT-UNet-192 \cite{xu2021levit} ~  &  $80.65\pm0.33$ & $91.85\pm0.06$  & $78.15\pm0.22$  & $88.15\pm0.11$ \\
% LeViT-UNet-384 \cite{xu2021levit} ~ & $82.95\pm0.13$ & $93.34\pm0.06$ & $81.71\pm0.20$ & $90.94\pm0.22$ \\

% \bottomrule
% \hline
% \end{tabular}
% \end{table*}

\noindent \textbf{Effectiveness of CTFT:} In Table \ref{Tab:ctft_ablation}, we show that adding CTFT to the base network improves the segmentation performance. To further Validate the claim, we conduct more experiments as seen in Table \ref{ctft}. It can be observed that adding CTFT to a joint-UNet architecture results in a  boost in performance.

\begin{table*}[!h]
\caption{Ablation Study. All results are reported in Dice score.}
\label{Tab:ctft_ablation}
\centering
\medskip
\begin{tabular}{@{}lcccccccccccc@{}}
\toprule
& \multicolumn{2}{c}{SonixTouch} & \multicolumn{2}{c}{Clarius} \\
\cmidrule(lr){2-3} \cmidrule(lr){4-5}
Method & Surface (\%) & Shadow (\%) & Surface (\%) & Shadow (\%) \\
\midrule

Joint-UNet ~ & $76.45\pm0.03$  & $86.06\pm0.15$ & $75.11\pm0.33$ & $84.01\pm0.17$ \\
Joint-UNet + CTFT ~ &  $77.19\pm 0.17$  & $89.01\pm0.15$ & $75.81\pm0.21$ & $84.71\pm0.11$ \\

\bottomrule
\hline
\end{tabular}
\label{ctft}
\end{table*}

\vspace{-1 em}
 
\section{Conclusion}

Accurate, complete, and robust bone and shadow segmentation are important to make ultrasound an essential imaging modality in clinically acceptable orthopedics procedures. In this paper, we propose an  end-to-end network to simultaneously perform robust and accurate bone and shadow segmentation by leveraging mutual information between the two tasks. The main novelty of our work lies in (1) the first systematic design of exploiting interrelation between two tasks to improve both bone and shadow segmentation, and (2) the design of fusion method of CNN and vision transformer to leverage multi-task learning while optimizing accuracy-efficiency trade-off. We believe the multi-task learning framework is an important contribution to the field of US-based orthopedic procedures.

\bibliographystyle{splncs04}
\bibliography{references}

\end{document}